\documentclass[preliminary,copyright]{eptcs}
\providecommand{\event}{IWIGP 2011} 

\usepackage{amsmath}
\usepackage{amsthm}
\usepackage{amssymb}
\usepackage{enumerate}
\usepackage{xspace}
\usepackage{breakurl}

\usepackage{tikz}
\usetikzlibrary{automata,patterns,decorations}
\usepgflibrary{shapes.multipart,plotmarks}

\usepackage[draft,english]{fixme}
\usepackage{xspace}

\newcommand{\dotcup}{\mathrel{\mathaccent\cdot\cup}}

\newcommand{\aut}[1]{\ensuremath{\mathcal{#1}}}
\newcommand{\autapprS}[1]{\ensuremath{\mathcal{#1}/_{\approx_S}}}

\newcommand{\G}{\ensuremath{\Gamma}\xspace}
\newcommand{\pz}{Player~0\xspace}
\newcommand{\po}{Player~1\xspace}

\newcommand{\rhds}{\ensuremath{\leq^{rh}_{de}}\xspace}
\newcommand{\rhde}{\ensuremath{\approx^{rh}_{de}}\xspace}

\newcommand{\Grhde}{\ensuremath{\mc{G}^{rh}_{de}}\xspace}
\newcommand{\Vrhde}{\ensuremath{V^{rh}_{de}}\xspace}
\newcommand{\Erhde}{\ensuremath{E^{rh}_{de}}\xspace}

\newcommand{\appr}{\ensuremath{\approx}\xspace}
\newcommand{\apprS}{\ensuremath{\approx_S}\xspace}
\newcommand{\al}{\ensuremath{\alpha}\xspace}

\newcommand{\LA}{\ensuremath{L(\aut{A})}\xspace}
\newcommand{\Nat}{\ensuremath{\mathbb{N}}\xspace}
\newcommand{\B}{\ensuremath{\mathbb{B}}\xspace}
\renewcommand{\S}{\ensuremath{\Sigma}\xspace}

\newcommand{\ie}{i.e.\@\xspace}
\newcommand{\cf}{cf.\@\xspace}

\newcommand{\wlofg}{w.l.o.g.\@\xspace}

\newcommand{\Inf}{\ensuremath{\mathrm{Inf}}}

\newcommand{\ld}{\ensuremath{\ldots}\xspace}
\newcommand{\cd}{\ensuremath{\cdots}\xspace}
\newcommand{\om}{\ensuremath{\omega}\xspace}
\newcommand{\D}{\ensuremath{\Delta}\xspace}
\renewcommand{\d}{\ensuremath{\delta}\xspace}

\newcommand{\qs}{\ensuremath{q_{\text{\scriptsize{sink}}}}\xspace}
\newcommand{\smin}{\ensuremath{s_{\min}}\xspace}
\newcommand{\Lra}{\ensuremath{\Longrightarrow}\xspace}

\newcommand{\RR}{Request-Response\xspace}
\newcommand{\rr}{request-response\xspace}
\newcommand{\M}{Muller\xspace}
\newcommand{\St}{Streett\xspace}

\newcommand{\es}{\ensuremath{\varnothing}\xspace}
\newcommand{\subseq}{\ensuremath{\subseteq}\xspace}
\newcommand{\vrho}{\ensuremath{\varrho}\xspace}
\newcommand{\sm}{\ensuremath{\setminus}\xspace}
\newcommand{\IAR}{\ensuremath{\mathrm{IAR}}\xspace}

\renewcommand{\t}{\ensuremath{\times}}
\newcommand{\StV}{\ensuremath{S\t V}\xspace}
\newcommand{\cl}[1]{\ensuremath{\mathrm{cl}(\aut{#1})}\xspace}
\newcommand{\Om}{\ensuremath{\Omega}\xspace}

\newcommand{\Recurz}{\ensuremath{\mathrm{Recur}_0}\xspace}
\newcommand{\cm}{\ensuremath{\checkmark}\xspace}

\renewcommand{\mod}{\ensuremath{\mathrm{mod}}\xspace}

\newcommand{\Dupl}{Duplicator\xspace}

\newcommand{\mc}[1]{\ensuremath{\mathcal{#1}}\xspace}
\renewcommand{\O}[1]{\ensuremath{\mc{O}(#1)}\xspace}
\renewcommand{\pm}{\ensuremath{\mathrm{pm}}\xspace}
\newcommand{\Sp}{\ensuremath{\mathrm{Sp}}\xspace}
\newcommand{\Du}{\ensuremath{\mathrm{Du}}\xspace}
\newcommand{\pInit}{\ensuremath{\mathrm{p_I}}\xspace}

\theoremstyle{definition}
\newtheorem{algorithm}{Algorithm}
\newtheorem{definition}{Definition}
\newtheorem{example}{Example}

\theoremstyle{remark}
\newtheorem{remark}{Remark}

\theoremstyle{plain}
\newtheorem{lemma}{Lemma}
\newtheorem{theorem}{Theorem}
\newtheorem{corollary}{Corollary}

\newif\iflong\longtrue

\title{Memory Reduction via Delayed Simulation}
\author{
Marcus Gelderie \qquad\qquad Michael Holtmann
\email{\{gelderie,holtmann\}@automata.rwth-aachen.de}
\institute{RWTH Aachen University, Lehrstuhl f\"ur Informatik 7, D-52056 Aachen}
}
\def\titlerunning{Memory Reduction via Delayed Simulation}
\def\authorrunning{Marcus Gelderie, Michael Holtmann}
\begin{document}
\maketitle

\begin{abstract}
We address a central (and classical) issue in the theory of infinite games: the reduction of the memory size that is needed to implement winning strategies in regular infinite games (\ie, controllers that ensure correct behavior against actions of the environment, when the specification is a regular \om-language). We propose an approach which attacks this problem before the construction of a strategy, by first reducing the game graph that is obtained from the specification. For the cases of specifications represented by ``request-response''-requirements and general ``fairness'' conditions, we show that an exponential gain in the size of memory is possible.
\end{abstract}

\section{Introduction}\label{sec:introduction}
Infinite games are a tool for the construction and verification of reactive systems. We consider the case of two players, \textit{Player~0} modeling a controller and \textit{Player~1} its environment. We deal with finite arenas and regular winning conditions, the latter one being captured by standard automata theoretic acceptance conditions (for example B\"uchi or \M conditions). For these types of games the winner is computable and a finite-state winning strategy can be constructed \cite{BL69SolSeqCondFinStateStr,GH82TreesAutGames,WAL04LandGamesBack,GTW02AutLogInfGam}.

There are many criteria for measuring the quality of a winning strategy. If only a finite memory is needed, then we mostly consider the size of this memory. This view has been pursued in many papers, among them \cite{DJW97HowMuchMem}. For example, it is known that weak and strong \M games over a graph with $n$ vertices can be solved with winning strategies of size at most $\O{2^n}$ and $\O{n!}$, respectively. There are well-known examples that show the optimality of these bounds \cite{DJW97HowMuchMem}.

A standard method for the construction of winning strategies is to proceed in two steps. In a first step, the game graph $G$ is expanded by a memory structure $S$, yielding a larger game graph $G'$ (loosely indicated as $G'=S\t G$), while the original winning condition $\varphi$ is transformed into a simpler one ($\varphi'$), allowing for positional winning strategies over $G'$. In the case of weak \M games the memory structure is the powerset of the set of all vertices of $G$, whereas in strong \M games we consider the set of all sequences of vertices of $G$. From a positional winning strategy over $G'$ one immediately obtains a finite-state winning strategy over $G$ with memory $S$. A reduction of the memory size can then be performed by classical minimization algorithms for sequential functions (as they are computed by Mealy automata).

In this paper we pursue an alternative approach that addresses the aspect of memory reduction at an earlier stage, namely \textit{before} the construction of a positional winning strategy over $G'$. More precisely, we insert an intermediate step of reducing $G'$, viewing it as an acceptor \aut{A} of an \om-language, where the winning condition $\varphi'$ is used as acceptance condition. This reduction yields a smaller graph $G'_0$ with memory structure $S_0$; it has the same type of winning condition as $G'$. For the graph $G'_0$ we construct a positional winning strategy, which is subsequently transformed into a finite-state winning strategy with memory structure $S_0$ over $G$.
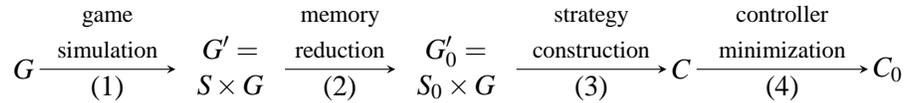
\begin{figure}
\begin{center}
	\begin{tikzpicture}[node distance=2.75cm,minimum width=0cm,minimum height=0cm,
						shorten <=2pt,shorten >=2pt,>=stealth,
						inner sep=0pt]
	\node (G) {$G$};					
	\node (StimesG)[node distance=2.75cm,right of=G] {\begin{tabular}{c}
														$G'=$\\
														$S\t G$
													\end{tabular}};
	\node (G0)[node distance=3cm,right of=StimesG] {\begin{tabular}{c}
															$G'_0=$\\
															$S_0\t G$
														\end{tabular}};
	\node (C1)[node distance=3cm,right of=G0] {$C$};
	\node (C01)[right of=C1] {$C_0$};
	\path[->] (G) edge node[above]{\begin{tabular}{c}
										\footnotesize{game}\\
										\footnotesize{simulation}
									\end{tabular}}
						node[below]{\begin{tabular}{c}
										($1$)
									\end{tabular}}(StimesG)
			(StimesG) edge node[above] {\begin{tabular}{c}
											\footnotesize{memory}\\
											\footnotesize{reduction}
										\end{tabular}}
							node[below]{\begin{tabular}{c}
											($2$)
										\end{tabular}}(G0)
			(G0) edge node[above] {\begin{tabular}{c}
										\footnotesize{strategy}\\
										\footnotesize{construction}
									\end{tabular}}
						node[below]{\begin{tabular}{c}
										($3$)
									\end{tabular}}(C1)
			(C1) edge node[above] {\begin{tabular}{c}
										\footnotesize{controller}\\
										\footnotesize{minimization}
									\end{tabular}}
						node[below]{\begin{tabular}{c}
										($4$)
									\end{tabular}}(C01);
	\end{tikzpicture}
\end{center}
\caption{Our Memory Reduction Approach (see Step 2)}
\end{figure}

The main challenge in this approach is to introduce a method for reducing \om-automata with acceptance conditions that are known to define games with positional winning strategies. State space reduction of \om-automata is a difficult problem, already for B\"uchi conditions. In \cite{HL07MemRed} our method was presented for the case of weak \M games. Reduction of weak parity automata can be done by the method of minimization presented in \cite{Loe01EffMin}. It turned out that the approach can result in an exponential gain, regarding the size of the memory needed.

In the present paper we address the more difficult case of ``strong'' winning conditions. Whereas a weak winning condition merely refers to visits and non-visits to vertices, a strong winning condition considers the set of vertices visited infinitely often. We deal with two particular winning conditions of practical interest, namely \RR and \St conditions.

In a \RR game the winning condition is a conjunction of statements ``whenever a state with property $p$ is visited, then sometime later a state with property $q$''. Formally, we are given a set $\Om=\{(P_1,R_1),\ld,(P_k,R_k)\}$ of pairs of subsets of $V$. \pz wins if every request, \ie, a visit to some $P_j$, is eventually followed by a matching response, \ie, the set $R_j$ is visited sometime later. A \St winning condition is denoted similarly, we call the sets $E_j,F_j$. \pz wins if infinitely many visits to $F_j$ imply infinitely many visits to $E_j$.

We propose a method to reduce the size of \om-automata with the aforementioned types of acceptance conditions, using this to reduce the size of game graphs before the construction of winning strategies. For the case of B\"uchi automata we apply the approach of ``delayed'' simulation presented in \cite{EWS01FairSim}. A state $q$ is delayed simulated by a state $q'$ if, in a run from $q$, each visit to a final state can eventually be answered by a visit to a final state in a run from $q'$. This condition is tested via a simulation game between two players. For the parity condition, as obtained by a game simulation of a \St game, we use an extended version of delayed simulation \cite{FW06SimRelPar}. If in the run from the simulated state a particular color is seen, then in the run from the simulating state this color has to be exceeded by a better one (regarding the acceptance condition). In our setting, computation of delayed simulation for the B\"uchi condition can be reduced to minimization of standard DFA, whereas for the parity condition we have to solve the corresponding simulation game explicitly.

This work is structured as follows: In the subsequent section we recall the basic terminology and known results. Section~\ref{sec:reduction_of_game_graphs} presents our approach in abstract terms. The main issue here is to reconcile the two views of a graph as used for the presentation of an infinite game and for the definition of an \om-language. Section~\ref{sec:request_response} develops the approach for the case of request-response games and shows an example where an exponential gain in the size of the memory needed for implementing a winning strategy is obtained. In Section~\ref{sec:Streett} we treat analogously the case of \St games.

\section{Preliminaries}\label{sec:preliminaries}
An \textit{infinite game $\G=(G,\varphi)$} is played by two players, \pz and \po. The \textit{game arena} is a finite directed graph $G=(V,E)$ with each vertex belonging to either player, \ie, $V=V_0\dotcup V_1$ where $V_0$ belongs to \pz and $V_1$ belongs to \po, and $E\subseq V\t V$. The \textit{winning condition} $\varphi\subseq V^\om$ is the set of all infinite paths through $G$ which are winning for \pz. Starting from an initial vertex the players move a token along edges in $E$, building up an infinite \textit{play $\vrho=\vrho(0)\vrho(1)\vrho(2)\cd$}. If the current vertex belongs to $V_0$, then \pz moves the token, and analogously for \po. The play $\vrho$ is \textit{winning for Player~0} if $\vrho\in\varphi$, otherwise it is winning for \po.

A \textit{strategy} for \pz is a function $f$ defining a next move for every game position of \pz (analogously for \po), \ie, it is a partial function $f:V^*V_0\to V$ such that for every play prefix $v_0\cd v_k$ with $v_k\in V_0$ it holds $(v_k,f(v_0\cd v_k))\in E$. A play $\vrho=\vrho(0)\vrho(1)\vrho(2)\cd$ is \textit{played according to $f$} if for all $\vrho(i)\in V_0$ it holds $\vrho(i+1)=f(\vrho(0)\cd\vrho(i))$. A strategy $f$ is called a \textit{winning strategy from $v$} for \pz if each play starting in $v$ that is played according to $f$ is winning for \pz. The \textit{winning region} $W_0$ of \pz is the set of all vertices from where \pz has a winning strategy. Strategies can be implemented by I/O-automata in the format of Mealy-automata.

In this work we deal with \textit{Request-Response} and \textit{Streett} games. A \rr winning condition is given by a set $\Om=\{(P_1,R_1),\ld,(P_k,R_k)\}$ of pairs of subsets of $V$. A request is a visit to a set $P_j$, and a response is a visit to a set $R_j$, for $1\leq j\leq k$. A play $\vrho=\vrho(0)\vrho(1)\vrho(2)\cd\in V^{\om}$ is winning for \pz if and only if every request is eventually responded to, \ie, for every $j$ it holds
\[
	\forall i(\vrho(i)\in P_j\Lra\exists i'\geq i:\vrho(i')\in R_j)
\]
A \St winning condition is induced by a set $\Om=\{(E_1,F_1),\ld,(E_k,F_k)\}$ of pairs of subsets of $V$. For a play $\vrho$, let $\Inf(\vrho)$ be the set of vertices visited infinitely often in $\vrho$. The play is winning for \pz if and only if for every pair $(E_j,F_j)$ it holds
\[
	\Inf(\vrho)\cap F_j\neq\es\Lra\Inf(\vrho)\cap E_j\neq\es.
\]
For the algorithm we are going to introduce we need the notion of \textit{game simulation}. The key idea of a game simulation is to extend the given game graph by a memory component such that on the new game graph a simpler winning condition can be used to simulate the original one. Any solution to the extended game can be used to compute an I/O-automaton that implements a winning strategy for the original game (see for example \cite{Tho95SynthStrInfGames}).
\begin{definition}\label{def:game_simulation}
Let $\G=(G,\varphi)$ and $\G'=(G',\varphi')$ be infinite games with game graphs $G=(V,E)$ and $G'=(V',E')$ and winning conditions $\varphi,\varphi'$. We say that \G is \textit{simulated} by $\G'$ (short: $\G\leq\G'$) if and only if the following hold:
\begin{enumerate}
	\item\label{item:definition_game_simulation_set_of_vertices} $V'=\StV$ for a finite memory set $S$ (and $(s,v)\in V'_i\iff v\in V_i$)
	\item\label{item:definition_game_simulation_play_transformation} There exists $s_0\in S$ such that every play $\vrho$ of \G is transformed into a unique play $\vrho'$ of $\G'$ by
		\begin{enumerate}
			\item\label{item:definition_game_simulation_play_transformation_initial_memory_content} $\vrho(0)=v$ $\Lra$ $\vrho'(0)=(s_0,v)$
			\item\label{item:definition_game_simulation_memory_update} Let $(s,v)\in V'$:
				\begin{enumerate}
					\item\label{item:definition_game_simulation_play_transformation_memory_update_existence} $(v,v')\in E$ $\Lra$ $\exists s'\in S:((s,v),(s',v'))\in E'$
					\item\label{item:definition_game_simulation_play_transformation_memory_update_uniqueness} $((s,v),(s_1,v_1))\in E',((s,v),(s_2,v_2))\in E'$ $\Lra$ $s_1=s_2$
				\end{enumerate}
			\item\label{item:definition_game_simulation_play_transformation_minimality} $((s,v),(s',v'))\in E'$ $\Lra$ $(v,v')\in E$
		\end{enumerate}
	\item\label{item:definition_game_simulation_winning_condition} $\vrho$ is winning for \pz in \G $\iff$ $\vrho'$ is winning for \pz in $\G'$
\end{enumerate}
\end{definition}
Later on, we present game simulation algorithms for \rr games by \textit{B\"uchi} games and for \St games by \textit{parity} games \cite{WHT03SymbSynth,GTW02AutLogInfGam}. A B\"uchi winning condition is induced by a set $F\subseq V$ of final vertices. A play $\vrho$ is winning for \pz if and only if the set $F$ is visited infinitely often: $F\cap\Inf(\vrho)\neq\es$. We solve a B\"uchi game as follows: First, we compute the set $\Recurz(F)\subseq F$ consisting of all final vertices from where \pz can force infinitely many visits to~$F$. Afterwards, we compute the $0$-Attractor of $\Recurz(F)$, \ie, the set of all vertices from where \pz can force a visit to~$\Recurz(F)$. It can be shown that this attractor coincides with the winning region of \pz in the B\"uchi game and that an associated attractor winning strategy reduces the distance to~$F$ in each move (\cf $\Pi^0_2$-games in~\cite{Tho95SynthStrInfGames}). A parity winning condition is given by a \textit{coloring} of the set of vertices, \ie, a function $c:V\to\{0,\ld,m\}$. A play $\vrho$ is winning for \pz if and only if the maximal color seen infinitely often, denoted $\max(c(\Inf(\vrho)))$, is even.

All types of games we consider are \textit{determined}: from each vertex one of the players has a winning strategy. In the sequel, we denote a B\"uchi game $(G,F)$ rather than $(G,\varphi)$, and analogously for other types of games. We assume that the reader is familiar with the basic theory of \om-automata (see for example \cite{Tho97LangAutLog,GTW02AutLogInfGam}).

\section{Reduction of Game Graphs}\label{sec:reduction_of_game_graphs}
The idea of our memory reduction algorithm is to reduce the game graph $G'$ before computing a winning strategy. To get this in a formal setting we transform infinite games into $\omega$-automata, and vice versa. We view the simulating game $\G'$ as an \om-automaton \aut{A} accepting exactly the plays winning for \pz in \G. The automaton \aut{B} is obtained from \aut{A} by state space reduction in such a way that the structural properties of game simulation are preserved, \ie, \G is simulated by $\G''$, where $\G''$ is the automaton \aut{B} viewed as infinite game. To reduce \aut{A} we compute a language-preserving equivalence relation on the memory $S$.
\begin{definition}\label{def:game_automaton}
Let $\G=(G,\varphi)$ and $\G'=(G',\varphi')$ be infinite games such that $\G\leq\G'$. We define the (deterministic) \textit{game automaton} $\aut{A}=((\StV)\dotcup\{q_0,\qs\},q_0,\d,\psi,V_0)$ over $V$. The function \d is adopted from $E'$ and a transition is labeled by the $V$-component of its target state. For $v'\in V$ we set $\d(q_0,v'):=(s_0,v')$ and $\d(\qs,v'):=\qs$. For $s\in S,v,v'\in V$ with $(v,v')\notin E$ we set $\d((s,v),v'):=\qs$. The acceptance condition $\psi$ is defined on an abstract level: A run $q_0\vrho'$ of \aut{A} is defined accepting if and only if $\vrho'$ is a winning play for \pz in $\G'$. (Conversely, an \textit{automaton game} is constructed from a game automaton in the obvious way. For that we need to keep $V_0$ in \aut{A}.)
\end{definition}
A simulating game and its game automaton are equivalent in the following sense.
\begin{remark}\label{rem:game_equivalent_to_game_automaton}
Let $\G,\G'$ be infinite games such that $\G\leq\G'$ and \aut{A} the game automaton of $\G'$. Then, \aut{A} accepts exactly the plays winning for \pz in \G: $\LA=\varphi$.
\end{remark}
We now reduce the game automaton in such a way that the properties of game simulation are preserved. To retain item~\ref{item:definition_game_simulation_set_of_vertices} from Definition~\ref{def:game_simulation} we compute an equivalence relation \appr on \StV and refine it to \apprS on $S$, where only \apprS is used for reduction. Moreover, to achieve item~\ref{item:definition_game_simulation_winning_condition} we have to preserve the language of \aut{A}. We require the following structural properties for \appr.
\begin{definition}\label{def:compatible}
Let \aut{A} be a game automaton and let \appr be an equivalence relation on \StV. We say that \appr is \textit{compatible} with \aut{A} if and only if the following hold:
\begin{center}
	\begin{enumerate}
		\item\label{item:definition_compatible_transition_structure} For all $s_1,s_2\in S,v,v'\in V$:\\
			  $(s_1,v)\appr(s_2,v)\Lra\d((s_1,v),v')\appr\d((s_2,v),v')$
		\item\label{item:definition_compatible_merged_runs} Let $\rho$ and $\rho'$ be two runs in \aut{A} (starting at arbitrary states) such that it holds $\rho(i)\appr\rho'(i)$, for all $i\in\Nat$. Then $\rho$ is accepting if and only if $\rho'$ is accepting.
	\end{enumerate}
\end{center}
\end{definition}
The quotient automaton of \aut{A} with respect to \apprS is defined on the basis of the following observation: If $(s_1,v)\appr(s_2,v)$ holds then from these two states exactly the same inputs are accepted. (Note that \aut{A} gets as inputs the plays of the game \G.) If this is true for all $v\in V$, then $s_1$ and $s_2$ can be considered equivalent.
\begin{definition}\label{def:apprS}
Let \aut{A} be a game automaton and let \appr be a compatible equivalence relation on \StV. The equivalence relation \apprS on $S$ is defined as follows:
\begin{center}
	$s_1\apprS s_2:\iff\forall v\in V:(s_1,v)\approx(s_2,v)$
\end{center}
\end{definition}
For $s\in S$, $[s]$ denotes the equivalence class of $s$ with respect to \apprS. Given $s_1\apprS s_2$ and $(v,v')\in E$, let $(s'_i,v'):=\d((s_i,v),v')$ for $i=1,2$. According to Definition~\ref{def:compatible} we know that $(s'_1,v')\approx(s'_2,v')$ holds. However, $s'_1\apprS s'_2$ does not hold necessarily. To get the $v'$-successor of $([s_1],v)$ well-defined we use some fixed total order $\prec_S$ on $S$.
\begin{definition}\label{def:apprS_game_automaton}
Let \appr be compatible and \apprS be derived from it as above. We define the \textit{quotient automaton} $\autapprS{A}=((S/_{\apprS}\t V)\dotcup\{q_0,\qs\},q_0,\d/_{\apprS}$,$\psi/_{\apprS},V_0)$ over $V$. Given $([s],v)\in S/_{\apprS}\t V$ and $(v,v')\in E$ we define
\begin{center}\label{pageref:smin}
	$\d/_{\apprS}(([s],v),v'):=([\smin],v')$
\end{center}
where
\begin{center}
	$\smin:=\min\{\hat{s}'\mid\exists\hat{s}:\hat{s}\approx_S s\text{ and }\d((\hat{s},v),v')=(\hat{s}',v')\}$.
\end{center}
The rest of $\d/_{\apprS}$ is defined analogously. Let $\rho=q_0([s_1],v_1)([s_2],v_2)\cd$ be a run of \autapprS{A}. We define $\rho$ to be accepting if and only if the run $\rho'=q_0(s'_1,v_1)(s'_2,v_2)\cd$ of \aut{A} (which is uniquely determined by $\rho$) is accepting.
\end{definition}
The run $\rho'$ is uniquely determined by $\rho$ because both \aut{A} and \autapprS{A} are deterministic. The acceptance condition for \autapprS{A} immediately implies $\LA=L(\autapprS{A})$. Later on, we show that for reducing B\"uchi game automata and parity game automata there exist compatible equivalence relations which are computable efficiently from a game automaton \aut{A}. Moreover, the respective quotient automaton \autapprS{A} can be defined with the same type of acceptance condition, in both cases. The following theorem shows that the automaton game $\G''$ of \autapprS{A} has the same structural properties as $\G'$, \ie, \G is simulated by $\G''$. (For the proof see \cite{HL07MemRed}.)
\begin{theorem}[\cite{HL07MemRed}]\label{thm:memory_reduction}
Let $\G=(G,\varphi)$ and $\G'=(G',\varphi')$ be infinite games such that $\G\leq\G'$. Let $\aut{A}$ be the game automaton of $\G'$ and \appr a compatible equivalence relation on \StV. Then $\G$ is simulated by the automaton game $\G''$ of \autapprS{A}.
\end{theorem}
We present the full algorithm for memory reduction.
\begin{algorithm}\label{alg:memory_reduction}({\sc{Memory Reduction}})\\
	Input: Infinite game $\G=(G,\varphi)$\\
	Output: Strategy automaton $\aut{A}_f$ for \pz from $W_0$
	\begin{enumerate}
		\item Establish a game simulation of \G by a new game $\G'$ in which \pz has a positional winning strategy from $W'_0$ (\cf Definition~\ref{def:game_simulation}).
		\item View $\G'$ as \om-automaton \aut{A} (\cf Definition~\ref{def:game_automaton}).
		\item\label{step:algorithm_memory_reduction} Reduce \om-automaton \aut{A}: Use a compatible equivalence relation \appr on \StV to compute \apprS on $S$ and construct the corresponding quotient game automaton \autapprS{A} (\cf Definitions~\ref{def:compatible},\ref{def:apprS_game_automaton}).
		\item View \autapprS{A} as automaton game $\G''=(G'',\varphi'')$ (\cf Definition~\ref{def:game_automaton}).
		\item Compute a positional winning strategy for \pz in $\G''$ and from it construct the strategy automaton $\aut{A}_f$.
	\end{enumerate}
\end{algorithm}
Algorithm~\ref{alg:memory_reduction} does not depend on the actual winning condition $\varphi$, but we need a suitable relation \appr to execute step~\ref{step:algorithm_memory_reduction}. Moreover, Theorem~\ref{thm:memory_reduction} is even valid if $\G'$ does not admit positional winning strategies.

\section{\RR Games}\label{sec:request_response}
In this section we apply the framework of the preceding section to \rr games. The first step is a game simulation by a B\"uchi game. The idea of this simulation is to memorize the set of open requests, and we use a marker (which is cyclically increased) to indicate which request is to be fulfilled next. Every time the marker is reset to value $1$ we visit a final state.
\begin{remark}\label{rem:game_simulation_request_response_by_Buechi}
Let $G=(V,E)$ be a game graph and $\Om=\{(P_1,R_1),\ld,(P_k,R_k)\}$ a family of $k$ pairs of subsets of $V$. Then the induced \RR game $\G=(G,\Om)$ is simulated by a B\"uchi game $\G'=(G',F')$.
\end{remark}
\begin{proof}
We define the game graph $G'=(V',E')$ and the set $F'$ of final vertices as follows:
\begin{itemize}
	\item[$\bullet$] $V':=2^{\{1,\ld,k\}}\t\{1,\ld,k\}\t\{0,1\}\t V$
	\item[$\bullet$] $((P,i,b,v),(P',i',b',v'))\in E':\iff$
		  \begin{itemize}
			\item[-] $P'=(P\cup\{i\mid v\in P_i\})\sm\{i\mid v\in R_i\}$
			\item[-] $i' = \left\{ \begin{array}{l@{\hspace{0.3cm}}l}
							i & \text{if }i\in P'\\
							(i~\mod~k)+1 & \text{otherwise}
						\end{array} \right. $
			\item[-] $b' = \left\{ \begin{array}{l@{\hspace{0.3cm}}l}
							1 & \text{if }i=k\text{ and }i'=1\\
							0 & \text{otherwise}
						\end{array} \right. $
			\item[-] $(v,v')\in E$
		  \end{itemize}
	\item[$\bullet$] $F':=2^{\{1,\ld,k\}}\t\{1,\ld,k\}\t\{1\}\t V$
\end{itemize}
\end{proof}
It is easy to verify (see \cite{WHT03SymbSynth}) that the above construction satisfies Definition~\ref{def:game_simulation}. In the sequel, we explain how to compute a compatible equivalence relation for B\"uchi game automata, using results on delayed simulation presented in \cite{EWS01FairSim}.

\subsection{Delayed Simulation for B\"uchi Automata}\label{subsec:delayed_simulation_Buechi}
The delayed simulation game $\mc{G}_{de}(q_0,q'_0)$ on a B\"uchi automaton \aut{A} is played by two players, \textit{Spoiler} and \textit{Duplicator}, and starts at $(q_0,q'_0)$, where $q_0,q'_0$ are arbitrary states of \aut{A}. In the first round Spoiler chooses a transition $(q_0,a_0,q_1)\in\D$ and Duplicator answers by a transition $(q'_0,a_0,q'_1)\in\D$ with the same labeling. From the pair $(q_1,q'_1)$ the game proceeds with the second round, analogously, and so on, until infinity. This way, Spoiler and Duplicator build up two infinite paths $\rho=q_0q_1q_2\cd$ and $\rho'=q'_0q'_1q'_2\cd$, respectively. The play $(\rho,\rho')$ is winning for Duplicator if and only if
\[
	\forall i(q_i\in F\Lra\exists j\geq i:q'_j\in F).
\]
We say that \textit{$q'_0$ delayed simulates $q_0$} if and only if Duplicator has a winning strategy in $\mc{G}_{de}(q_0,q'_0)$ and denote this $q_0\preceq_{de}q'_0$. Moreover, we say that $q_0,q'_0$ delayed simulate each other, denoted $q_0\simeq_{de}q'_0$, if and only if $q_0\preceq_{de}q'_0$ and $q'_0\preceq_{de}q_0$. Quotienting with respect to $\simeq_{de}$ preserves the recognized language.
\begin{lemma}[\cite{EWS01FairSim}]\label{lem:delayed_simulation_is_language_preserving}
Let \aut{A} be a B\"uchi automaton. Then it holds $\LA=L(\aut{A}/_{\simeq_{de}})$.
\end{lemma}
We extend delayed simulation to delayed \textit{bisimulation} as follows: In each round of the play, Spoiler has a free choice whether to take the next transition in either $\rho$ or $\rho'$, and Duplicator must take the next transition in the other run, afterwards. The winning condition is modified as follows: If a final state is seen at position $i$ of either run, then there must exist $j\geq i$ such that a final state is seen at position $j$ in the other run. If Duplicator wins the corresponding game, then we say that $q_0,q'_0$ are delayed \textit{bisimilar}, and denote this $q_0\appr_{de}q'_0$.

We make use of the fact that, for deterministic B\"uchi automata, delayed simulation can be replaced by delayed bisimulation, and vice versa.
\begin{remark}\label{rem:delayed_simulation_and_delayed_bisimulation_coincide}
Let \aut{A} be a deterministic B\"uchi automaton. Then, for all states $q,q'$ of \aut{A} it holds
\[
	q\simeq_{de}q'\iff q\appr_{de}q'.
\]
\end{remark}
Remark~\ref{rem:delayed_simulation_and_delayed_bisimulation_coincide} follows immediately from the fact that the transitions taken in \aut{A} are uniquely determined by the letters chosen by Spoiler. Hence, it makes no difference whether he chooses the next transition in $\rho$ or $\rho'$. To compute $\appr_{de}$ we introduce \textit{direct} bisimulation. It is defined as delayed bisimulation with the only difference of a modified winning condition in the corresponding game: The play $(q_0q_1q_2\cd,q'_0q'_1q'_2\cd)$ is winning for Duplicator if and only if
\[
	\forall i(q_i\in F\iff q'_i\in F).
\]
If Duplicator has a winning strategy in the direct bisimulation game, starting at $(q_0,q'_0)$, then we say that $q_0,q'_0$ are direct bisimilar, and we denote this $q_0\appr_{di}q'_0$. To explain how delayed bisimulation and direct bisimulation are connected, we introduce the \textit{closure}\label{pageref:closure_operator} of \aut{A}, denoted \cl{A}. It has the set $F'$ of final states, where we initially set $F':=F$ and iterate the following until a fixed point is reached:
\begin{center}
	If there exists $q\notin F'$ such that all successors of $q$ are in $F'$, then put $q$ in $F'$.
\end{center}
The automata \aut{A} and \cl{A} are equivalent and, clearly, \cl{A} is deterministic if \aut{A} is deterministic. Moreover, note that \cl{A} can be computed in time linear in $|\aut{A}|$. The following lemma completes the approach to state space reduction of (deterministic) B\"uchi automata.
\begin{lemma}[\cite{EWS01FairSim}]\label{lem:delayed_bisimulation_on_A_and_direct_bisimulation_on_closure_of_A_coincide}
Let \aut{A} be a B\"uchi automaton. For all states $q,q'$ we have
\[
	q\appr_{de}q'\text{ in }\aut{A}\iff q\appr_{di}q'\text{ in }\cl{A}.
\]
\end{lemma}
For a given B\"uchi game automaton \aut{A} we compute \autapprS{A} as follows: We compute the direct bisimulation relation $\appr_{di}$ in \cl{A} which coincides with $\simeq_{de}$ in \aut{A}. Note that for a deterministic B\"uchi automaton the computation of $\appr_{di}$ is the same as block partitioning for standard DFA. Hence, the computation can be done in time $\mc{O}(n\log{n})$, if $n$ is the number of states of \cl{A} and $|\S|$ is assumed constant \cite{PT87PartRef}. As a direct consequence, we get that $\appr_{di}$ is compatible with \cl{A} and $\appr_{de}$ (hence, also $\simeq_{de}$) is compatible with \aut{A}. From $\simeq_{de}$, the relation \apprS is computed as given in Definition~\ref{def:apprS} where \appr is replaced by $\simeq_{de}$. Finally, we can apply Theorem~\ref{thm:memory_reduction}.
\begin{remark}\label{rem:compatible_delayed_bisimulation}
Let \aut{A} be a B\"uchi game automaton and $\simeq_{de}$ the delayed simulation relation for \aut{A}. Then $\simeq_{de}$ is compatible with \aut{A}.
\end{remark}
\begin{corollary}\label{cor:memory_reduction_request_response}
Let \G be a \RR game and $\G'$ the corresponding B\"uchi game (\cf Remark~\ref{rem:game_simulation_request_response_by_Buechi}). Further, let \aut{A} be the game automaton of $\G'$ and $\simeq_{de}$ defined as above. Then \G is simulated by the automaton game $\G''$ of \autapprS{A}.
\end{corollary}
Let us briefly give the memory reduction algorithm for \RR games.
\begin{algorithm}{({\sc{Memory Reduction for \RR games}})}\label{alg:memory_reduction_request_response}\\
	Input: \RR game $\G=(G,\Om)$\\
	Output: Strategy automaton $\aut{A}_f$ for \pz from $W_0$
	\begin{enumerate}
		\item Establish a game simulation of \G by a B\"uchi game $\G'$ (\cf Remark~\ref{rem:game_simulation_request_response_by_Buechi}).
		\item View $\G'$ as B\"uchi game automaton \aut{A} (\cf Definition~\ref{def:game_automaton}).
		\item Compute \cl{A} (\cf page~\pageref{pageref:closure_operator}) and the direct bisimulation relation $\appr_{di}$ in \cl{A}. By Lemma~\ref{lem:delayed_bisimulation_on_A_and_direct_bisimulation_on_closure_of_A_coincide} and Remark~\ref{rem:delayed_simulation_and_delayed_bisimulation_coincide} it coincides with $\simeq_{de}$ in \aut{A}. From $\simeq_{de}$ compute \apprS (\cf Definition~\ref{def:apprS}).
		\item View \autapprS{A} as B\"uchi automaton game $\G''$ (\cf Definition~\ref{def:game_automaton}).
		\item Compute a positional winning strategy for \pz in $\G''$ and from it construct $\aut{A}_f$.
	\end{enumerate}
\end{algorithm}
We have shown that the delayed simulation relation of a deterministic B\"uchi automaton can be computed in time \O{n\cdot\log{n}}, where $n$ is the number of states and $|\S|$ is assumed constant. Here, we get a complexity of \O{n\cdot(\log{n})^2}, because we have \O{\log{n}} input letters. The B\"uchi game $\G''$ can be solved in time \O{n^2\cdot\log{n}}. Hence, the total running time of Algorithm~\ref{alg:memory_reduction_request_response} is polynomial in $|\G'|$.

\subsection{An Example for \RR Games}\label{subsec:rr_example}
In this subsection we compare the memory size of winning strategies (obtained by Algorithm~\ref{alg:memory_reduction_request_response}) with the standard approach, where after the conversion of a \rr game into a B\"uchi game a winning strategy is directly computed (and then possibly minimized according to I/O-automata minimization). We present an example with an exponential gain in the memory size.

Consider the game graph $G_k$, which is shown in Figure~\ref{fig:request_response_exponential_to_positional} for $k=3$. Let \Om be the following \rr winning condition:
\[
	\Om_k=\{(P_0,R_0)\}\cup\{(P_1,R_1),(P'_1,R'_1),\ld,(P_k,R_k),(P'_k,R'_k)\}
\]
A play proceeds as follows: From the initial vertex $v$, \po takes $k$ decisions activating either $P_i$ or $P'_i$ ($i=1,\ld,k$). At vertex $w$ \pz takes over making $k$ decisions himself. In vertex $y$ all pairs from \Om are responded to. Hence, each (positional) strategy for \pz is winning.
\begin{figure}[ht]
	\begin{center}	
\begin{tikzpicture}[node distance=1.75cm,minimum width=0.5cm,minimum height=0.5cm,
					shorten <=2pt,shorten >=2pt,>=stealth,
					inner sep=0pt]
	\node[draw] (1) {$v$};
	\node (P_0label) [node distance=0.5cm,above of=1] {$P_0$};
	
	\node[draw] (P_1) [node distance=0.875cm,right of=1,yshift=0.875cm] {};
	\node (P_1label) [node distance=0.5cm,above of=P_1] {$P_1$};
	\node[draw] (P'_1) [below of=P_1] {};
	\node (P'_1label) [node distance=0.5cm,below of=P'_1] {$P'_1$};

	\node[draw] (2) [right of=1] {};

	\node[draw] (P_2) [right of=P_1] {};
	\node (P_2label) [node distance=0.5cm,above of=P_2] {$P_2$};
	\node[draw] (P'_2) [right of=P'_1] {};
	\node (P'_2label) [node distance=0.5cm,below of=P'_2] {$P'_2$};
	
	\node[draw] (3) [right of=2] {};
	
 	\node[draw] (P_3) [node distance=1.75cm,right of=P_2] {};
 	\node (P_3label) [node distance=0.5cm,above of=P_3] {$P_3$};
 	\node[draw] (P'_3) [node distance=1.75cm,right of=P'_2] {};
 	\node (P'_3label) [node distance=0.5cm,below of=P'_3] {$P'_3$};
	
	\node[draw,circle] (1') [node distance=0.875cm,right of=P_3,yshift=-0.875cm] {$w$};
	\node (R_0label) [node distance=0.5cm,above of=1'] {$R_0$};
	
	\node[draw] (R_1) [node distance=0.875cm,right of=1',yshift=0.875cm] {};
	\node (R_1label) [node distance=0.5cm,above of=R_1] {$R_1$};
	\node[draw] (R'_1) [below of=R_1] {};
	\node (R'_1label) [node distance=0.5cm,below of=R'_1] {$R'_1$};
	
	\node[draw,circle] (2') [right of=1'] {};
	
	\node[draw] (R_2) [right of=R_1] {};
	\node (R_2label) [node distance=0.5cm,above of=R_2] {$R_2$};
	\node[draw] (R'_2) [right of=R'_1] {};
	\node (R'_2label) [node distance=0.5cm,below of=R'_2] {$R'_2$};
	
\node[draw,circle] (3') [right of=2'] {};
	
 	\node[draw] (R_3) [node distance=1.75cm,right of=R_2] {};
 	\node (R_3label) [node distance=0.5cm,above of=R_3] {$R_3$};
 	\node[draw] (R'_3) [node distance=1.75cm,right of=R'_2] {};
 	\node (R'_3label) [node distance=0.5cm,below of=R'_3] {$R'_3$};

	\node[draw] (x) [node distance=0.875cm,right of=R_3,yshift=-0.875cm] {$x$};
	\node[draw] (y) [node distance=1cm,right of=x] {$y$};

	\node (ydummy)[node distance=1.25cm,above of=y,xshift=-0.4cm] {$\textstyle{\forall i:R_i,R'_i}$};

	\path[->] (1) edge (P_1)
			  (1) edge (P'_1)
			  (P_1) edge (2)
			  (P'_1) edge (2)
			  
			  (2) edge (P_2)
			  (2) edge (P'_2)
			  
(P_2) edge (3)			  
(P'_2) edge (3)
(3) edge (P_3)
(3) edge (P'_3)

			  (P_3) edge (1')
			  (P'_3) edge (1')

			  (1') edge (R_1)
			  (1') edge (R'_1)
			  (R_1) edge (2')
			  (R'_1) edge (2')
			  
			  (2') edge (R_2)
			  (2') edge (R'_2)
(R_2) edge (3')
(R'_2) edge (3')

(3') edge (R_3)
(3') edge (R'_3)
			  (R_3) edge (x)
			  (R'_3) edge (x)
			  
			  
(x) edge (y)
(y) edge[loop below] ()
(ydummy) edge[dotted,bend left=15] (y)
;
\end{tikzpicture}
		\caption{\label{fig:request_response_exponential_to_positional} \textsl{\RR Game Graph $G_3$}}
	\end{center}
\end{figure}
\begin{theorem}\label{thm:request_response_exponential_to_positional}
Let $\G_k=(G_k,\Om_k)$ be the \RR game from Figure~\ref{fig:request_response_exponential_to_positional} and let $\G'_k=(G'_k,F')$ be the B\"uchi game simulating $\G_k$, constructed as in the proof of Remark~\ref{rem:game_simulation_request_response_by_Buechi}. Then, \pz wins $\G_k$ from $v$ such that the following hold:
\begin{enumerate}
	\item The positional winning strategy $f'_k$ for \pz from $(\es,1,0,v)$ in $\G'_k$ yields a winning strategy $f_k$ for \pz from $v$ in $\G_k$ of size at least $2^k$.
	\item The reduced game graph $G''_k$ computed by Algorithm~\ref{alg:memory_reduction_request_response} has only one memory content.
\end{enumerate}
\end{theorem}
\begin{proof}
If \pz precisely mimics the decisions of \po, \ie, for all $i=1,\ld,k$ she moves to the $R_i$-vertex if and only if \po has moved to the $P_i$-vertex before, then in the B\"uchi game a final vertex is seen as soon as vertex $y$ is visited for the first time. On the way from vertex $w$ to vertex $y$, the counter in the second component of the memory is increased by one for $2k+1$ times: It starts with value $1$ at vertex $w$ and has value $2k+1$ when reaching vertex $x$; it is reset to value $1$ when the play proceeds from vertex $x$ to vertex $y$.

If \pz makes a mistake, \ie, there exists $i$ such that she moves to the $R_i$-vertex if and only if \po has moved to the $P'_i$-vertex, then a final vertex is reached several moves later than in the case where she plays ``correctly''. This is due to the fact that her false decision at the $i$th response avoids that the counter in the second component of the memory is increased.

By our remarks above, there is a unique shortest path from vertex $w$ to vertex $y$ which visits a final vertex in the B\"uchi game. It is the path which precisely mimics the path from vertex $v$ to vertex $w$. Solving the B\"uchi game we obtain an attractor strategy, which means that a final vertex is assumed as soon as possible. Hence, the strategy chooses the ``correct'' path from vertex $w$ to vertex $y$, and this strategy requires a memory of size at least $2^k$ because it needs to memorize each of the $k$ decisions of \po.

Now, let us consider Algorithm~\ref{alg:memory_reduction_request_response}. Let $\G'$ be the B\"uchi game computed by the game simulation from Remark~\ref{rem:game_simulation_request_response_by_Buechi}, and let $s:=(\es,1,1,y)\in\StV$. Once the play on $G$ reaches vertex $y$, the set of active pairs is emptied, whereby on~$G'$ we reach a vertex of the form~$(\es,i,b,y)$, where $1\leq i\leq k,b\in\B$. After at most $k$ revisits to vertex $y$ (on $G$) the value of $i$ is reset to $1$, which means that we reach vertex $s$ (on $G'$). Moreover, vertex $s$ is repeatedly visited every $2k+1$ moves, thereafter. Summing up, every infinite path on $G'$ visits vertex $s$ (infinitely often).

Consider the B\"uchi game automaton \aut{A} of $\G'$. By the remarks above, every infinite path in \aut{A} (without \qs) leads through $s$, even when starting at non-reachable states. Accordingly, all states in \StV are declared final in \cl{A}. Thus, we obtain $(s_1,v)\appr_{di}(s_2,v)$ for all $s_1,s_2\in S,v\in V$. Accordingly, all memory contents are equivalent, \ie $S/_{\apprS}$ is a singleton.
\end{proof}

\section{\St Games}\label{sec:Streett}
A \St winning condition is given by a family \Om of pairs of subsets of $V$:
\[
	\Om=\{(E_1,F_1),\ld,(E_k,F_k)\}
\]
\pz wins if and only if, for each $j$, infinitely many visits to $F_j$ imply infinitely many visits to $E_j$.

In a game simulation for \St games by parity games we keep track of the order of the latest visits to the sets $F_i,E_i$ ($i=1,\ld,k$). To do so, we use a data structure called \textit{Index Appearance Record}, short \IAR \cite{GTW02AutLogInfGam}. For $k\geq1$, we denote $\mc{S}_k$ the symmetric group of $\{1,\ld,k\}$, \ie, the set of all its permutations.
\begin{remark}\label{rem:game_simulation_Streett_by_parity}
Let $G=(V,E)$ be a game graph and $\Om=\{(E_1,F_1),\ld,(E_k,F_k)\}$ a family of pairs of subsets of $V$. Then the induced \St game $\G=(G,\Om)$ is simulated by a parity game $\G'=(G',c')$.
\end{remark}
\begin{proof}
Let $G'=(V',E')$ be defined as follows. As memory $S$ we use the Index Appearance Record (\IAR) of $V$:
\begin{center}
	$S:=\IAR(V)=\{(i_1\cd i_k,e,f)\mid(i_1\cd i_k)\in\mc{S}_k,1\leq e,f\leq k\}$
\end{center}
As initial memory content we choose $s_0:=(1\cd k,1,1)$. The transition relation $E'$ is uniquely determined by $E$ and $\Om$. We define:\\
$(((i_1\cd i_k,e,f),v),((i'_1\cd i'_k,e',f'),v'))\in E':\iff$
\begin{enumerate}
	\item $(v,v')\in E$
	\item $(i'_1\cd i'_k)$ is obtained from $(i_1\cd i_k)$ by shifting all $i_l$ with $v\in E_{i_l}$ to the left, $l\in\{1,\ld,k\}$
	\item\label{pageref:game_simulation_Streett_by_parity_pointer_e_prime} $e'$ is the maximal\footnote{We assume \wlofg that $E_k=F_k=V$ to have the pointers $e'$ and $f'$ well-defined.} $l\in\{1,\ld,k\}$ such that $v\in E_{i_l}$
	\item $f'$ is the maximal $m\in\{1,\ld,k\}$ such that $v\in F_{i'_m}$
\end{enumerate}
The coloring $c':\IAR(V)\t V\to\{1,\ld,2k\}$ of $V'$ is defined by:
\[ c'((i_1\cd i_k,e,f),v) := \left\{ \begin{array}{l@{\hspace{0.3cm}}l}
					2e & \text{if }e\geq f\\
					2f-1 & \text{if }e<f\\
				\end{array} \right. \]
\end{proof}
We use the right-hand delayed simulation for alternating parity automata (introduced in \cite{FW06SimRelPar}) to reduce parity game automata. Whereas for B\"uchi game automata the problem of computing delayed simulation can be reduced to the minimization problem for standard DFA (\cf Section~\ref{subsec:delayed_simulation_Buechi}), for parity game automata we have to solve the corresponding simulation game explicitly. It is described in \cite{FW06SimRelPar}, and we can use a simplified version of it. Firstly, a parity game automaton is not alternating which means that the first move of each round is made by Spoiler in the simulated automaton and the second move is made by Duplicator in the simulating automaton. Due to this fixed order of moving the pebbles we need less vertices in the simulation game graph. Secondly, a parity game automaton is deterministic. This means that the positions of the two pebbles and the update of the priority memory (see below) are uniquely determined by the letter chosen by Spoiler. Hence Duplicator's moves are predetermined by Spoiler's moves and, accordingly, all vertices in the simulation game graph belong to Spoiler. Let us define the simulation game in a formal way.

\subsection{Right-hand Delayed Simulation for Parity Automata}\label{subsec:delayed_simulation_parity}
We are given a parity game automaton $\aut{A}=((\StV)\dotcup\{q_0,\qs\},q_0,\d,c',V_0)$ over $V$, where a run $\rho$ of \aut{A} is accepting if and only if the maximal color seen infinitely often in $\rho$ is even. In \cite{FW06SimRelPar} a $\min$-parity condition is assumed. Hence, we have to redefine the coloring of \aut{A} by $c':=k-c'$ for even $k\in\Nat$ large enough. We construct the simulation game $\Grhde=(G^{rh}_{de},\varphi^{rh}_{de})$ as follows: The game graph $G^{rh}_{de}=(\Vrhde,\Erhde)$ has the set of vertices $\Vrhde=(\StV)\t(\StV)\t(c'(\StV)\dotcup\{\cm\})$, where $c'(\StV)$ denotes the set of colors assigned by the parity function $c'$. We set $V_{\Sp}:=\Vrhde$ (and $V_{\Du}:=\es$). The edge relation $\Erhde\subseq\Vrhde\t\Vrhde$ is defined as follows:
\begin{center}
	$(((s_1,v_1),(s_2,v_2),k),((s'_1,v'_1),(s'_2,v'_2),k'))\in\Erhde:\iff$\\
	$((s_1,v_1),(s'_1,v'_1))\in E',((s_2,v_2),(s'_2,v'_2))\in E',v'_1=v'_2$ and\\
	$k'=\pm(c'(s'_1,v'_1),c'(s'_2,v'_2),k)$
\end{center}
The \textit{priority memory} update function $\pm:\Nat\t\Nat\t(\Nat\dotcup\{\cm\})\to\Nat\dotcup\{\cm\}$ is defined as follows:
\begin{enumerate}[i.]
	\item\label{case:definition_priority_memory_case_one} $\pm(i,j,\cm)=\min\{i,j\}$, if $i\prec j$
	\item $\pm(i,j,\cm)=\cm$, if $j\preceq i$
	\item\label{case:definition_priority_memory_case_three} $\pm(i,j,k)=\min\{i,j,k\}$, if $i\prec j$
	\item\label{case:definition_pm_improved_version}\label{case:definition_priority_memory_case_four} $\pm(i,j,k)=k$, if $j\preceq i$, $i$ is odd and $i\leq k$, and $j$ is odd or $k<j$
	\item\label{case:definition_priority_memory_case_five} $\pm(i,j,k)=\cm$, if $j\preceq i$, $j$ is even and $j\leq k$, and $i$ is even or $k<i$
	\item\label{case:definition_priority_memory_case_six} $\pm(i,j,k)=\cm$, if $i$ is odd, $j$ is even, and both $i\leq k$ and $j\leq k$
	\item\label{case:definition_priority_memory_else}\label{case:definition_priority_memory_case_seven} else $\pm(i,j,k)=k$
\end{enumerate}
In the basic definition of the delayed simulation game in \cite{FW06SimRelPar} the value of \pm in case~\ref{case:definition_pm_improved_version} is set to \cm. It is also shown there that quotienting with respect to the obtained equivalence relation is not language-preserving. Hence, we use the slightly modified version from above, where in case~\ref{case:definition_pm_improved_version} the value of \pm is set to $k$. The induced relation is defined on page~\pageref{pageref:definition_right_hand_delayed_simulation} and preserves the recognized language. (This is also shown in \cite{FW06SimRelPar}.) The binary relation $\prec$ is the \textit{reward order}\label{pageref:reward_order_on_natural_numbers} on \Nat. For $m,n\in\Nat$, we define $m\preceq n$ if and only if
\begin{enumerate}
	\item $m$ is even and $n$ is odd, or
	\item $m$ and $n$ are both even and $m\leq n$, or
	\item $m$ and $n$ are both odd and $n\leq m$.
\end{enumerate}
This yields $0\prec2\prec4\prec\ld\prec5\prec3\prec1$. If $m\prec n$ then we say that $m$ is \textit{better} than $n$, whereas terms like \textit{minimum} and \textit{smaller} refer to the standard relation $<$ on \Nat. We\label{pageref:explanation_definition_priority_memory_else} leave it up to the reader to verify that case~\ref{case:definition_priority_memory_else} of the definition of \pm applies if and only if $j\preceq i,k<i$ and $k<j$.

A play $\vrho$ is winning for Duplicator if and only if the set
\[
	F:=(\StV)\t(\StV)\t\{\cm\}
\]
is visited infinitely often; this means that $\varphi^{rh}_{de}$ is a B\"uchi condition. (Spoiler wins if and only if he can avoid \cm from a certain point onwards.) We\label{pageref:definition_right_hand_delayed_simulation} say that \textit{$(s_2,v_2)$ right-hand delayed simulates $(s_1,v_1)$}, denoted $(s_1,v_1)\rhds(s_2,v_2)$, if and only if Duplicator has a winning strategy in \Grhde from the initial game position $\pInit((s_1,v_1),(s_2,v_2))$ defined as follows. Let $i:=c'(s_1,v_1)$ and $j:=c'(s_2,v_2)$:
\[\pInit((s_1,v_1),(s_2,v_2)) := \left\{ \begin{array}{l@{\hspace{0.3cm}}l}
						((s_1,v_1),(s_2,v_2),\min\{i,j\}) & \text{if }i\prec j\\
						((s_1,v_1),(s_2,v_2),\cm) & \text{otherwise}\\
\end{array} \right. \]
In \cite{FW06SimRelPar} it is shown that \rhds is a preorder implying language containment, \ie, if $(s_1,v_1)\rhds(s_2,v_2)$ then $L(\aut{A}_{(s_1,v_1)})\subseq L(\aut{A}_{(s_2,v_2)})$. We define the corresponding equivalence relation \rhde as
\[
	(s_1,v_1)\rhde(s_2,v_2):\iff(s_1,v_1)\rhds(s_2,v_2)\text{ and }(s_2,v_2)\rhds(s_1,v_1).
\]
Duplicator's winning region $W_{\Du}$ in \Grhde determines \rhds, and from that we can compute \rhde. Note that we need to consider only the case where it holds $v_1=v_2$ (\cf Definition~\ref{def:apprS}).

\subsection{Quotienting}\label{subsec:quotienting}
The relation \rhde is compatible with a parity game automaton. Item~\ref{item:definition_compatible_transition_structure} of Definition~\ref{def:compatible} is verified by the upcoming lemma. Item~\ref{item:definition_compatible_merged_runs} of Definition~\ref{def:compatible} follows from the fact that quotienting with respect to \rhde is language-preserving, as is shown in \cite{FW06SimRelPar}.
\begin{lemma}\label{lem:compatible_rhde}
Let~\aut{A} be a parity game automaton and \rhde defined as in Section~\ref{subsec:delayed_simulation_parity}. Then, for all $s_1,s_2\in S,v_1,v_2,v'\in V$ it holds:
\[
	(s_1,v_1)\rhde(s_2,v_2)\Lra\d((s_1,v_1),v')\rhde\d((s_2,v_2),v')
\]
\end{lemma}
\begin{remark}\label{rem:compatible_right_hand_delayed_simulation}
Let~\aut{A} be a parity game automaton and \rhde the right-hand delayed simulation relation for~\aut{A}. Then \rhde is compatible with \aut{A}. The quotient automaton $\aut{A}/_{\rhde}$ is defined in the natural way: $\d/_{\rhde}([(s,v)],v'):=[\d((s,v),v')]$ and $c'/_{\rhde}([(s,v)]):=\min\{c'(s',v')\mid(s',v')\rhde(s,v)\}$; $\aut{A}/_{\rhde}$ is equivalent to \aut{A}. We refer to \cite{FW06SimRelPar} for the details.
\end{remark}
We compute \apprS from \rhde (as in Definition~\ref{def:apprS}) and express the acceptance condition $\psi/_{\apprS}$ of \autapprS{A} in terms of a coloring $c'/_{\apprS}$. To this end, let $s\in S,v\in V$ and define
\[
	c'/_{\apprS}([s],v):=\min\{c'(s',v')\mid(s',v')\rhde(s,v)\},
\]
and let $q_0,\qs$ inherit their color from \aut{A}. Since $s_1\apprS s_2$ implies $(s_1,v)\rhde(s_2,v)$ for all $v\in V$, the above definition of $c'/_{\apprS}$ is independent of representatives. Note that \autapprS{A} is a game automaton. Essentially, the relation \apprS is a refinement of \rhde. Hence, the automaton \autapprS{A} is equivalent to \aut{A}.
\begin{lemma}\label{lem:apprS_game_automaton_equivalent}
Let \aut{A} be a parity game automaton and \autapprS{A} the corresponding \apprS-quotient with coloring $c'/_{\apprS}$ (see above). Then \aut{A} and \autapprS{A} are equivalent.
\end{lemma}
\begin{proof}
We have to show $\LA=L(\autapprS{A})$, where it suffices to show $L(\aut{A}/_{\rhde})=L(\autapprS{A})$. By Lemma~\ref{lem:compatible_rhde} automaton $\aut{A}/_{\rhde}$ is deterministic, and by Definition~\ref{def:apprS_game_automaton} automaton \autapprS{A} is deterministic. For $\al\in V^\om$, let $\rho$ be the run of \autapprS{A} on \al and $\rho'$ be the corresponding run of $\aut{A}/_{\rhde}$ on \al. The run $\rho'$ is uniquely determined by the run $\rho$, because both $\aut{A}/_{\rhde}$ and \autapprS{A} are deterministic and \apprS is a refinement of $\rhde$. Moreover, $\rho$ is accepting if and only if $\rho'$ is accepting, because both runs have the same sequence of colors. Since both \autapprS{A} and $\aut{A}/_{\rhde}$ are deterministic, there is no other run on \al, neither for \autapprS{A} nor for $\aut{A}/_{\rhde}$. Thus, \al is accepted by \autapprS{A} if and only if it is accepted by $\aut{A}/_{\rhde}$.
\end{proof}
Our above results show that our algorithm for memory reduction is applicable to a \St game \G as follows: We simulate \G by a parity game $\G'$ which is then transformed into a parity game automaton \aut{A}. For \aut{A} we construct the right-hand delayed simulation game \Grhde and solve it by standard techniques \cite{GTW02AutLogInfGam}. Duplicator's winning region in this game and Definition~\ref{def:apprS} uniquely determine \apprS. The corresponding quotient automaton \autapprS{A} is a parity game automaton equivalent to \aut{A}, and we can transform it into a unique parity automaton game $\G''$. By Theorem~\ref{thm:memory_reduction}, \G is simulated by $\G''$.
\begin{corollary}\label{cor:memory_reduction_Streett}
Let \G be a \St game and $\G'$ the corresponding parity game (\cf Remark~\ref{rem:game_simulation_Streett_by_parity}). Further, let \aut{A} be the game automaton of $\G'$ and \rhde defined as above. Then \G is simulated by the automaton game $\G''$ of \autapprS{A}.
\end{corollary}
This yields the following algorithm.
\begin{algorithm}{({\sc{Memory Reduction for \St games}})}\label{alg:memory_reduction_Streett}\\
	Input: \St game $\G=(G,\Om)$\\
	Output: Strategy automaton $\aut{A}_f$ for \pz from $W_0$
	\begin{enumerate}
		\item\label{step:memory_reduction_Streett_game_simulation} Establish a game simulation of \G by a parity game $\G'$ (\cf Remark~\ref{rem:game_simulation_Streett_by_parity}).
		\item\label{step:memory_reduction_Streett_game_automaton} View $\G'$ as parity game automaton \aut{A} (\cf Definition~\ref{def:game_automaton}); redefine the coloring of \aut{A} as $c':=2k-c'$.
		\item\label{step:memory_reduction_Streett_compute_delayed_simulation} Construct the delayed simulation game \Grhde for \aut{A} and solve it. From \Dupl's winning region compute \apprS (\cf Definition~\ref{def:apprS}).
		\item View \autapprS{A} as parity automaton game $\G''$ (\cf Definition~\ref{def:game_automaton}).
		\item Compute a positional winning\footnote{Note that $\G''$ is a $\min$-parity game.} strategy for \pz in $\G''$ and from it construct $\aut{A}_f$.
	\end{enumerate}
\end{algorithm}
At this point we have to mention an optional normalization, which may make the relation \rhde larger. It is done before executing step~\ref{step:memory_reduction_Streett_compute_delayed_simulation}. For each SCC $C$ of \aut{A} we iterate the following: While there exists $(s,v)\in C$ such that $c'(s,v)\geq2$ and there exists no $(s',v')\in C$ such that $c'(s',v')=c'(s,v)-1$ do $c'(s,v):=c'(s,v)-2$. Clearly, this does not change the accepted language \cite{FW06SimRelPar}.

For the computation of \rhde we need to solve the simulation game \Grhde. It is a B\"uchi game of size $\mc{O}(r^2\cdot k)$ where $r$ is the number of states of \aut{A}, and $k$ is the number of colors assigned by $c'$. Since B\"uchi games are solvable in polynomial time measured in the size of the game graph, the overall running time of Algorithm~\ref{alg:memory_reduction_Streett} is polynomial in $|\G'|$.

Note that the above technique can analogously be applied to \M games (see \cite{GH82TreesAutGames}). The only difference is the game simulation in step~\ref{step:memory_reduction_Streett_game_simulation}. \M games can also be simulated by parity games, but the needed memory depends on the number of vertices of the game graph $G$. Accordingly, we need to redefine $c':=2|V|-c'$ at the end of step~\ref{step:memory_reduction_Streett_game_automaton}. The rest of the algorithm is the same as for \St games, and we obtain a similar running time.

\subsection{An Example for \St Games}\label{subsec:Streett_example}
Let us show a result for the class of strong winning conditions, similar to the one above. We consider the \St game in Example~\ref{ex:Streett_exponential_to_positional} (see below) and make particular assumptions on the winning strategy for \pz in the simulating parity game. More precisely, we demand that she behaves ``optimal''. This is meant in the sense that she continuously chooses those edges which globally guarantee the best colors she can enforce. (A color $m$ is better than a color $n$ if $n\prec m$, where $\prec$ is the reward order from page~\pageref{pageref:reward_order_on_natural_numbers}.)
\begin{example}\label{ex:Streett_exponential_to_positional}
Let $G_k$ be the graph shown in Figure~\ref{fig:Streett_exponential_to_positional} (for $k=3$), and $\Om_k$ the following \St winning condition:
\[
	\Om_k=\{(E_1,F_1),(E_{-1},F_{-1}),\ld,(E_k,F_k),(E_{-k},F_{-k}),(V,V)\}
\]
\begin{figure}[ht]
	\begin{center}
\begin{tikzpicture}[node distance=1.75cm,minimum width=0.5cm,minimum height=0.5cm,
					shorten <=2pt,shorten >=2pt,>=stealth,
					inner sep=0pt]
	\node[draw] (1) {$v_1$};
	\node (arrow_anchor_A) [minimum width=0cm,minimum height=0cm,node distance=2cm,below of=1] {};
	
	\node[draw] (F_1) [node distance=0.875cm,right of=1,yshift=0.875cm] {};
	\node (F_1label) [node distance=0.5cm,minimum height=0cm,above of=F_1] {$E_{-1},F_1$};
	\node[draw] (F_-1) [below of=F_1] {};
	\node (F_-1label) [node distance=0.5cm,below of=F_-1] {$E_1,F_{-1}$};

	\node[draw] (2) [right of=1] {$v_2$};

	\node[draw] (F_2) [right of=F_1] {};
	\node (F_2label) [node distance=0.5cm,minimum height=0cm,above of=F_2] {$E_{-2},F_2$};
	\node[draw] (F_-2) [right of=F_-1] {};
	\node (F_-2label) [node distance=0.5cm,below of=F_-2] {$E_2,F_{-2}$};
	
	\node[draw] (3) [right of=2] {$v_3$};
	
 	\node[draw] (F_3) [node distance=1.75cm,right of=F_2] {};
 	\node (F_3label) [node distance=0.5cm,minimum height=0cm,above of=F_3] {$E_{-3},F_3$};
 	\node[draw] (F_-3) [node distance=1.75cm,right of=F_-2] {};
 	\node (F_-3label) [node distance=0.5cm,below of=F_-3] {$E_3,F_{-3}$};
	
	\node[draw,circle] (1') [node distance=0.875cm,right of=F_3,yshift=-0.875cm] {$w_1$};
	
	\node[draw] (E_1) [node distance=0.875cm,right of=1',yshift=0.875cm] {};
	\node (E_1label) [node distance=0.5cm,minimum height=0cm,above of=E_1] {$E_1,F_{-1}$};
	\node[draw] (E_-1) [below of=E_1] {};
	\node (E_-1label) [node distance=0.5cm,below of=E_-1] {$E_{-1},F_1$};
	
	\node[draw,circle] (2') [right of=1'] {$w_2$};
	
	\node[draw] (E_2) [right of=E_1] {};
	\node (E_2label) [node distance=0.5cm,minimum height=0cm,above of=E_2] {$E_2,F_{-2}$};
	\node[draw] (E_-2) [right of=E_-1] {};
	\node (E_-2label) [node distance=0.5cm,below of=E_-2] {$E_{-2},F_2$};
	
	\node[draw,circle] (3') [right of=2'] {$w_3$};
	
 	\node[draw] (E_3) [node distance=1.75cm,right of=E_2] {};
 	\node (E_3label) [node distance=0.5cm,minimum height=0cm,above of=E_3] {$E_3,F_{-3}$};
 	\node[draw] (E_-3) [node distance=1.75cm,right of=E_-2] {};
 	\node (E_-3label) [node distance=0.5cm,below of=E_-3] {$E_{-3},F_3$};

	\node[draw] (x) [node distance=0.875cm,right of=E_3,yshift=-0.875cm] {$x$};
	\node[draw] (y) [node distance=1cm,right of=x] {$y$};
	\node (arrow_anchor_B) [minimum width=0cm,minimum height=0cm,node distance=2cm,below of=y] {};

	\node (ydummy)[node distance=1.4cm,minimum width=0cm,above of=y,xshift=-0.1cm] {$\textstyle{\forall i:E_i,E_{-i}}$};

	\path[->] (1) edge (F_1)
			  (1) edge (F_-1)
			  (F_1) edge (2)
			  (F_-1) edge (2)
			  
			  (2) edge (F_2)
			  (2) edge (F_-2)
			  
(F_2) edge (3)			  
(F_-2) edge (3)
(3) edge (F_3)
(3) edge (F_-3)

			  (F_3) edge (1')
			  (F_-3) edge (1')

			  (1') edge (E_1)
			  (1') edge (E_-1)
			  (E_1) edge (2')
			  (E_-1) edge (2')
			  
			  (2') edge (E_2)
			  (2') edge (E_-2)
(E_2) edge (3')
(E_-2) edge (3')

(3') edge (E_3)
(3') edge (E_-3)
			  (E_3) edge (x)
			  (E_-3) edge (x)
			  
			  
(x) edge (y)
(ydummy) edge[dotted,bend left=15] (y)

(y) edge[-,shorten >=0pt] (arrow_anchor_B)
(arrow_anchor_B) edge[-,shorten <=0pt,shorten >=0pt] (arrow_anchor_A)
(arrow_anchor_A) edge[shorten <=0pt] (1)
;
\end{tikzpicture}
		\caption{\label{fig:Streett_exponential_to_positional} \textsl{\St Game Graph $G_3$}}
	\end{center}
\end{figure}

The game proceeds similarly to the one from Figure~\ref{fig:request_response_exponential_to_positional}. The major difference is that vertex~$v_1$ is visited infinitely often, naturally dividing each play into rounds. At the end of each round, \ie, when the play proceeds from vertex~$y$ to vertex~$v_1$, the highest possible color $4k+2$ is seen in the parity game $\G'_k$ (simulating $\G_k$). This is due to the fact that some index must be at the last position of the current \IAR and, accordingly, the pointer $e'$ has the value $2k+1$ (\cf page~\pageref{pageref:game_simulation_Streett_by_parity_pointer_e_prime}). Thus, each play satisfies the parity winning condition, and each (positional) strategy for \pz is winning.
\end{example}
\begin{theorem}\label{thm:Streett_exponential_to_positional}
Let $\G_k=(G_k,\Om_k)$ be the \St game from Example~\ref{ex:Streett_exponential_to_positional} and let $\G'_k=(G'_k,c_k)$ be the parity game simulating $\G_k$ (according to Remark~\ref{rem:game_simulation_Streett_by_parity}), where $s_0:=((1\cd2k+1),1,1)$ is the initial memory content. Then, \pz wins $\G_k$ from vertex~$v_1$ such that the following hold:
\begin{enumerate}
	\item\label{item:theorem_Streett_exponential_to_positional_complicated_winning_strategy} Each positional winning strategy $f'_k$ for \pz in $\G'_k$ from $(s_0,v_1)$ with
	\[
		\{c_k(s',v')\mid((s,v),(s',v'))\in E_{f'_k}\footnote{$E_{f'_k}$ denotes the set of all edges determined by the positional strategy $f'_k$.}\}\cap\{2n+1\mid n\in\Nat\}=\es
	\]
	yields a winning strategy $f_k$ for \pz in $\G_k$ from $v_1$ of size at least $2^k$.
%
%
	\item\label{item:theorem_Streett_exponential_to_positional_reduced_game_graph} The reduced game graph $G''_k$ computed by Algorithm~\ref{alg:memory_reduction_Streett} has only one memory content.
\end{enumerate}
\end{theorem}
\begin{proof}
For simplicity, we assume $k=3$; the proof is analogous for other values of $k$. First, we fix a convention on the entries in an \IAR: Let the pair $(V,V)$ be represented by $V$ and every other pair by its unique index, \ie, $(E_{-2},F_{-2})$ is represented by $\mathrel{-2}$; if index~$i$ has value $-j$, then let $\mathrel{-i:=j}$, for $1\leq j\leq3$.

The intersection in item~\ref{item:theorem_Streett_exponential_to_positional_complicated_winning_strategy} means that \pz chooses only edges leading into vertices of even color. To prove that item, note that the permutation reached (in $\G'_3$) when vertex~$w_1$ is reached (in $\G_3$) is of the form
\[
	(V\:i_3\cd i_1\:p),
\]
where $i_j\in\{j,-j\}$ (for $1\leq j\leq3$), and $p$ is some permutation of the set $\{-i_1,-i_2,-i_3\}$. For example, let \po at $v_1$ decide to move up, then again up, and then down; then the permutation is $(V\:\:3\mathrel{-2}\:\mathrel{-1}p)$, because the indices $-1$, $-2$ and $3$ are shifted to the second position one after another, and $V$ stays at the front (\cf proof of Remark~\ref{rem:game_simulation_Streett_by_parity}). Moreover, $p$ is a permutation of $\{1,2,-3\}$, \ie, of the set of all indices $i$ for which \po has moved to~$F_i$, recently.

If \pz moves upwards at~$w_1$, \ie, she mimics \po's behavior at vertex~$v_1$, then the permutation shortly becomes $(V\:\:1\:\:3\mathrel{-2}\:\mathrel{-1}p')$, and $p'$ is either $(2\mathrel{-3})$ or $(-3\:\:2)$. That means $e'$ is assigned the value $5$, $6$ or $7$.\footnote{Note that the definition of $e'$ refers to the old permutation, and index~$1$ came from position $5$, $6$ or $7$ (\cf page~\pageref{pageref:game_simulation_Streett_by_parity_pointer_e_prime}).} Simultaneously, $f'$ gets the value $5$ because $F_{-1}$ is visited and $-1$ is at the fifth position in the new permutation. Accordingly, it holds $e'\geq f'$, which means that we see color $10$, $12$ or $14$.

Conversely, if \pz moves downwards at~$w_1$, then the permutation becomes $(V\mathrel{-1}\:3\mathrel{-2}p'')$, with $p''=p$. Hence, $e'$ is assigned $4$ because index~$-1$ comes from the fourth position. Moreover, index~$1$ is located somewhere in $p''$, \ie, at position $5$, $6$ or $7$. Thus, we have $f'\geq5>4=e'$; accordingly, we see an odd color, either $9$, $11$ or $13$.

Making an analogous observation at vertices~$w_2,w_3$, we can deduce the following: If \pz mimics \po's behavior then she visits an even color, say $l$; if she makes the ``wrong'' move then a vertex of odd color less than $l$ is seen. Let the latter situation be called an \textit{error} and note that \pz can play errorless by memorizing \po's decisions. By an argument analogous to that in the proof of Theorem~\ref{thm:request_response_exponential_to_positional}, implementation of an errorless winning strategy requires a memory of size at least $2^k$.

To see item~\ref{item:theorem_Streett_exponential_to_positional_reduced_game_graph}, let us consider Algorithm~\ref{alg:memory_reduction_Streett}. First, note that the coloring $c_k$ is redefined as $c_k:=4k+2-c_k$ (in step~\ref{step:memory_reduction_Streett_game_automaton}); accordingly, we are dealing with a $\min$-parity condition from now on. Every play on $G'_k$ must traverse an edge $((s_1,y),(s_2,v))$ infinitely often, for some \IAR{s} $s_1,s_2$. Thereby, a vertex with the smallest possible color $0$ is visited, because there must be $1\leq j\leq k$ such that index~$j$ or~$-j$ is at the last position of the \IAR $s_1$. Thus, in the simulation game the priority memory is reset to \cm and a final vertex is visited, infinitely often. Accordingly, \Dupl has a winning strategy from each vertex in the simulation game graph. Summing up, all states (having the same $V$-component) in the parity game automaton \aut{A} of $\G'_k$ are \rhde-equivalent, which means that all memory contents are declared \apprS-equivalent. Thus, we obtain a reduced memory of size one.
\end{proof}

\section{Conclusion}\label{sec:conclusion}
We have presented a method that reduces the memory for implementing winning strategies in \RR and \St games. The key idea is to view the result of a game simulation as an $\omega$-automaton whose state space contains the memory to solve the given game. This state space is reduced via the notion of delayed simulation (\cf \cite{EWS01FairSim,FW06SimRelPar}). The reduction is carried out only on the set of memory contents, where two memory contents are considered equivalent if, from them, \pz wins exactly the same plays. In our setting, delayed simulation can be computed in time \O{n\cdot(\log{n})^2} and \O{n^2\cdot k} for B\"uchi and parity game automata, respectively, where $n$ is the number of states of the game automaton and $k$ the number of colors (in the parity game automaton). In both cases our algorithm has a running time polynomial in the size of the simulating game $\G'$.

\section*{Acknowledgment}
The authors thank Wolfgang Thomas for his advice.

\bibliographystyle{eptcs}
\bibliography{holtmann_bibliography}

\end{document}